# Is the new citation-rank approach P100' in bibliometrics really new?

Michael Schreiber

*Institute of Physics, Chemnitz University of Technology, 09107 Chemnitz, Germany.*
*Phone: +49 371 531 21910, Fax: +49 371 531 21919*
*E-mail: schreiber@physik.tu-chemnitz.de*

The percentile-based rating scale P100 describes the citation impact in terms of the distribution of unique citation values. This approach has recently been refined by considering also the frequency of papers with the same citation counts. Here I compare the resulting P100' with P100 for an empirical dataset and a simple fictitious model dataset. It is shown that P100' is not much different from standard percentile-based ratings in terms of citation frequencies. A new indicator P100" is introduced.

Keywords: evaluation, citation analysis, highly cited publications, bibliometric rankings, percentiles

### 1. Introduction

Citation distributions are usually extremely skewed with many uncited and lowly cited papers and a long tail with very few highly cited papers. This can create problems already for simple statistical analysis, like the interpretation of median or mean values. Recently, Bornmann, Leydesdorff, & Wang (2013) proposed a new approach based on the unique citation values which occur in a dataset. These citation frequencies are then attributed to percentiles between 0 and 100. Thus many papers with equal citation counts contribute only one value to the distribution for the new rating scale P100 so that a large reduction of the data is achieved (Bornmann & Mutz, 2014). A ranking in terms of P100 is therefore based on citation ranks.

One problem with the new approach is due to the long high-frequency citation tail of the distribution, where citation counts are often based on only one or two papers. This leads to reliability problems (Bornmann & Mutz, 2014), because small changes due to random fluctuations or due to the time evolution of the citation record can create or delete a unique citation value from the distribution and thus alter all percentile values (except the lowest and highest) leading to changes in the ranking. Corresponding counterintuitive effects in the performance evaluation in terms of P100 have been exemplified for two simple fictitious models and three larger empirical samples (Schreiber, 2014).

A similar paradoxical situation has been analyzed by Bornmann & Mutz (2014). It has led the authors to refine P100 by considering in the ranking also the frequency of papers with the same citation counts. However, the behavior of the resulting indicator P100' is completely different from P100. Moreover, it will be shown below that P100' is very similar to traditional percentile-based rating scales in terms of usual citation distributions.

In the following I shall compare P100 with P100' and P100' with traditional percentile-based indicators.

### 2. The P100 and P100' values for three protagonists of the new indicators

For the determination of the P100 indicator one has to determine the citation counts which occur in a reference set, for example all papers published for a certain subject category in a given year. The rank 100 is attributed to the highest citation count in the reference set and the rank 0 to the lowest citation count which usually will be zero. Sorting all unique citation counts $u_i$ is ascending order, i.e. $u_i < u_{i+1}$ from $i = 0$ to $i_{\max}$ allows one to attribute the ranks

$$P100(i) = 100\ i/i_{\max}. \qquad (1)$$

In my previous analysis (Schreiber, 2014) I discussed the values of P100 for the articles of the three senior authors who have proposed the P100 indicator. The results are presented in Fig. 1, where 270 papers with 4581 citations and 56 unique citation counts have been taken into account. The large impact of many papers by these authors is corroborated by high P100 values.

For the determination of the refined indicator

$$P100'(j) = 100\, j/j_{max} \qquad (2)$$

one has to consider also the frequency $n_i$ of the tied papers with the same citation count $u_i$. To be specific, the $n_0$ uncited papers get the rank $j = j_0 = 0$, the $n_1$ singly cited papers get the rank $j = j_1 = n_0$, the $n_2$ doubly cited papers are attributed to the rank $j = j_2 = j_1 + n_1 = n_0 + n_1$, and so on. In other words, all papers with $u_i$ citations obtain the same rank

$$j_i = j_{i-1} + n_{j-1} = \sum_{i'=0}^{i-1} n_{i'}. \qquad (3)$$

Thus the rank $j$ in (2) implicitly depends on the rank $i$ from (1). For the highest citation count one gets the highest rank

$$j_{max} = \sum_{i'=0}^{i_{max}-1} n_{i'} = N - n_{i_{max}} \qquad (4)$$

where $N$ is the total number of papers. Usually there is only one paper with the highest number of citations so that $j_{max} = N-1$. The subsequent discussion will assume that this is the case. Exceptions from this situation are analyzed in Section 4.

In Fig. 2 the values of P100' for the 270 papers of the three senior protagonists of P100 are visualized. A comparison with Fig. 1 shows that the data points lie significantly higher. This is a consequence of the strongly skewed citation distribution which leads to a large spreading of the $j_i$ values for low citation frequencies $u_i$ while for high citation frequencies $u_i$ there are usually very few papers so that the values of $j_i$ are narrowly spaced. Accordingly the transformation from P100 to P100' strongly shifts the results to higher values of the indicator. Thus Fig. 2 demonstrates much more clearly than Fig. 1 that many papers of these authors have a very large impact in comparison with the corresponding reference sets.

### 3. P100 and P100' in comparison with other percentile-based indicators

In order to analyze the behavior of P100 and P100' in more detail, I consider a specific reference set, namely all articles published in 2012 in the subject category "Information Science Library Science". The citation record was downloaded from Thomson Reuters' Web of Knowledge on 21 July 2014. An overview is given in Table 1. I have chosen a rather recent publication year, because then not so many unique citation counts occur which makes the table reasonably short. On the other hand this means that there are many uncited and lowly cited papers as can be seen in Table 1 where the 34 unique citation counts $u_i$ and the frequencies $n_i$ with which these citation counts occur in the reference set are shown. Altogether there are $N = 3424$ articles in this set with 6260 citations. The ranking in terms of unique citation counts and the resulting values of the P100 indicator are also given. One can see that 25 citations are sufficient to reach the top quartile. Only 9 papers thus make it into the top quartile. 27 papers have received more than 16 citations and are thus above the median of P100 scale.

The ranks $j_i$ which are relevant for the P100' indicator and the values of this indicator are also given in Table 1. The computation according to (3) thus proceeds from the bottom of the table upwards. As already seen by the comparison of Figs. 1 and 2, the high P100' values are much denser, the 30 papers with more than 15 citations fall into the top-1% category, and there are 287 papers with more than 5 citations qualifying for the top-10% class.

The InCites evaluation tool provided by Thomson Reuters' Web of Science sorts the papers by descending citation counts and then accumulates the number $N_i$ of papers with $u_i$ and more citations. Therefore, here the computation proceeds from the top of the table downwards. The results which are given in Table 1 are then normalized by the total number $N = N_0$ of papers (and multiplied by 100) yielding the percentiles given in Table 1. These are sometimes called inverted percentiles, because they are derived from *descending* citation counts. In order to make them comparable with the P100' scale, one has to invert the scale, substracting the values from 100. One can see in Table 1 that the thus derived inverted percentiles are very close and always slightly below the P100' indicator values. (Of course, there is no difference for the uncited papers, which are attributed to the zeroth percentile in both cases.) In fact, the calculation of the inverted InCites scale PiIC is equivalent to (2), replacing in the denominator $j_{max}$ by $N$ what means

$$\text{PiIC}(j_i) = 100\, j_i/N \qquad (5)$$

where the index $i$ makes the dependence of $j$ on the rank $i$ explicit. According to (4) this replacement is a minor change, because $n_{i_{max}} \ll N$. In the present reference set we have $n_{i_{max}} = 1$ as usual. But this means that the InCites scale starts with $n_{i_{max}} > 0$ so that the percentile 0 does not exactly occur. Correspondingly, the top percentile on the inverted scale will never by exactly equal to 100. That might be considered a disadvantage of this approach.

Leydesdorff and Bornmann (2011) proposed the "counting rule that the number of items with the lower citation rates than the item under study determines the percentile". All "tied, that is, precisely equal, numbers of citations thus are not counted as fewer than" (Leydesdorff, Bornmann, Mutz, & Opthof, 2011). These paper counts are indeed identical to the values of $j$ in (3) and thus given in Table 1 already for the calculation of P100'. But applying the mentioned counting rule for the determination of percentile values the normalization is again performed with the total number of papers $N$, not with $j_{max}$. Thus the resulting percentiles are the same as on the inverted InCites scale, i.e. they are given by (5).

If one includes the item under study into the number of items to compare with (Rousseau, 2012) one would expect only a difference of $100/N$ % which can be considered negligible for large reference sets. However, in this approach tied papers are all assigned to the largest percentile value. The effect is that the percentiles are shifted downwards by one line in Table 1 in comparison with the inverted InCites results. But the numerical values are exactly the same, what means

$$\text{PRou}(j_i) = 100\, j_{i+1}/N \qquad (6)$$

with

$$j_{i_{max}+1} = \sum_{i'=0}^{i_{max}} n_{i'} = N \qquad (7)$$

according to (3). Now the lowest percentile 0 cannot occur, but the highest percentile is always given by the value 100. It should be noted that the resulting values are always above the P100 indicator. The difference is very large for low citation frequencies, but small for high citation counts and of course vanishes for the most cited paper.

In my view the discussed rules can be considered as the extreme possibilities for the attribution of percentile values. Leydesdorff (2012) has interpreted these percentiles as lower and upper boundary of an uncertainty interval. It can be utilized for the fractional attribution of papers to different percentile rank classes (Schreiber, 2013, Waltman & Schreiber, 2013).

A compromise between the extreme possibilities is to utilize the middle of that interval (Leydesdorff, 2012) to categories the publications. Another approach was suggested by Pudovkin and Garfield (2009) who first attribute a percentile value between 1/$N$ and 1 to the papers without caring about equal citation counts. These are then taken into account by assigning the average of the percentiles of the tied publications to all the tied publications for a given citation frequency. The results are also given in Table 1. The deviation from the middle of the uncertainty interval always amounts to exactly 50/$N$ % = 0.013%, which is certainly negligible. The percentiles obtained by this average or by the middle of the uncertainty interval are very close to the P100' values around the middle of the citation distribution in terms of paper counts, i.e. when the accumulated number $N_i$ or $j$ of papers is around $N/2$. In Table 1 this cannot be observed, because there are so many singly cited papers.

In conclusion, the P100' values are always within the boundaries of the discussed uncertainty interval. This means that P100' is just another way of interpolating between the boundaries. Therefore in my view P100' is not a qualitatively new indicator. It does, however, have the advantage that the extreme values of P100' = 0 and P100' = 100 are utilized for the lowest and highest citation counts, respectively.

**4. Surprising behavior of P100'**

In the previous section I have demonstrated that P100' behaves very similar to traditional percentile-based approaches. There is, however, a situation in which surprising deviations can occur. This happens when more than one paper in the reference set has received the same highest number of citations. As an example, I consider a fictitious dataset with 5 papers and 4 unique citation counts as given in Table 2. It is not necessary to specify the actual citation frequencies for the subsequent discussion. But for the ease of the formulation I assume that the lowest citation count is zero. In any case, the P100 values are determined by $i_{max} = 3$ and shall not be changed. Initially there are two uncited papers so that the P100' values in Table 2 are determined by $j_{max} = 4$. They fall into the uncertainty intervals introduced in the previous section.

If one of the uncited papers receives so many citations that it reaches the second or third unique citation count, then the P100' values do not show an unexpected behavior, see also Table 2. But if it receives further citations and thus reaches the highest citation count, then $j_{max} = 3$ what leads to new values of P100'($j$) (see Table 3) and in particular to a value of P100'(2) = 67 which lies outside the uncertainty interval. I find this surprising. In this special case the behavior of P100' does not fulfill the expectations which I have had after the analysis in the previous section.

If there are more papers tied at the highest citation count, the effect is even stronger. This can be seen in Table 3 where I have also included the cases with 3 and 4 papers at the top level. We still have $j_{max} = 3$, therefore the P100' values are not influenced, but the uncertainty intervals are, because $N$ increases. In the last case (the fifth modification), not only P100'(2) but also P100'(3) drops out of the uncertainty intervals. Admittedly, such a citation distribution is contrary to the usual skewed behavior. The problem is therefore a rather artificial one and I expect that it does not occur in realistic empirical reference sets.

A related problem occurs when the second highest citation count disappears. This is demonstrated in Table 4. Here originally there are 5 papers with 5 different citation counts. Again the actual citation frequencies do not have to be specified. If the paper with the second highest citation count receives so many more citations that it draws level with most cited one, we have only 4 unique citation counts anymore in Table 4. Consequently, the P100' values for the lowly cited papers increase which is a counterintuitive behavior: Why should these papers be evaluated better than before? On the contrary, one would expect that their P100' values should rather decrease when another paper in the reference set improves its citation count. This paradoxical situation is reminescent of similar behaviors found for P100 (Schreiber, 2014) and was termed a reliability problem by Bornmann & Mutz (2014). But I consider it to be only a small deficiency, because it occurs only when there are tied papers at the top level of citations.

Another small deficiency which is inherent in all above discussed percentile-based indicators except P100 is the fact that usually the expectation value, i.e., the mean of the P100' values in the reference set does not amount to 50%. This is in contrast to the advantages of the citation-rank approach listed by Bornmann & Mutz (2014). P100 does indeed fulfill this requirement, but P100' does not, as can be easily seen for the initial situation and the second modification in Table 2 as well as for the second reference set discussed by Bornmann & Mutz (2014) in their table 2. If one does not average the P100' values directly, but rather utilizes the respective numbers of papers as weights which means effectively that one averages the P100' values of the papers instead of the citation levels, then *all* examples in Tables 2, 3, and 4 yield average values which are not equal to the middle of the percentile scale, i.e., not equal to 50%.

### 5. Summary, conclusion, and another new indicator P100"

The analysis of the P100' indicator for the citation record of all articles from 2012 in the subject category "Information Science Library Science" from the Web of Knowledge has shown, that the indicator values fall into the uncertainty interval of percentile values, as introduced by Leydesdorff (2012). The respective data from Table 1 are visualized in Fig. 3, where one can clearly see that P100 is completely different from P100' and that P100' is close to the inverted percentiles for low citation frequencies. For high citation frequencies the data points are so dense, that for distinction it is necessary to enlarge this part of the plot in Fig. 4. Now one can see that the P100' values coincide nearly with Rousseau's percentiles, if there is only $n_i = 1$ paper at the respective unique citation count $u_i$. If there are two papers with the same unique citation frequency, then P100' lies more or less in the middle of the uncertainty interval, otherwise it is closer to the lower boundary. This corroborates my conclusion of Section 3, namely that P100' is nothing but one more way of selecting a position within the uncertainty interval.

Unfortunately, as demonstrated in Section 4 this is not always the case: if there is more than one paper which has received the same highest number of citations, then deviations from the uncertainty interval can occur. Although this is an unusual situation, because usually there is only one paper at the top of the citation distribution, the effect is deplorable. Intuitively, I think that there should be a combination of P100 and P100' which avoids this problem. However, I have not been able to construct such a combination, which keeps the strong advantage from P100 and P100', namely that the value 0 is attributed to the lowest cited (usually the uncited) papers and that the value 100 is attributed to the most cited papers.

But an interpolation between PiIC and PRou by means of P100 does fulfill the expectations, namely

$$\begin{aligned} P100"(j_i) &= \text{PiIC}(j_i) + (\text{PRou}(j_i) - \text{PiIC}(j_i)) * i/i_{max} \\ &= 100\, j_i/N + (100\, j_{i+1}/N - 100\, j_i/N) * i/i_{max} \\ &= 100\, j_i/N + 100\, n_i/N * i/i_{max} \\ &= \text{PiIC}(j_i) + n_i/N * P100(i). \end{aligned} \quad (8)$$

The resulting values of the new indicator P100" are displayed in Figs. 3 and 4; they are given in Table 5 for the initial situation and the five modifications in Tables 2 and 3.

I am not sure whether I should seriously propose this construct as an alternative indicator. It is too complicated for my taste, but it shows that it is possible to avoid the problems described in the previous section: By its very construction P100" always lies within the uncertainty interval. This is corroborated by comparing Table 5 with Tables 2 and 3. Moreover, it can also be seen that for $i = 1$ and 2 the values of P100" decrease when the overall performance is improved, as it happens from modification to modification.

### Acknowledgement

I am grateful to Lutz Bornmann (Max Planck Society, Munich) for providing the values of the P100 and P100' indicators, which were utilized in Section 2. The values are from a bibliometrics database developed and maintained by the Max Planck Digital Library (MPDL, Munich) and derived from the Science Citation Index Expanded (SCI-E), Social Sciences Citation Index (SSCI), Arts and Humanities Citation Index (AHCI) prepared by Thomson Reuters Scientific, Philadelphia.

**Table 1.** Citation record of all articles from 2012 in the subject category "Information Science Library Science" with the unique citation counts $u_i$, the corresponding number of articles $n_i$ for rank $i$ and the resulting P100 values. The rank $j$ determines the P100' values. The cumulative number $N_i$ of papers yields the percentiles utilized by the InCites evaluation tool. The inverted InCites percentiles PiIC are also given, together with the percentiles PRou obtained from the rule of Rousseau (2012) and the average PPaG proposed by Pudovkin and Garfield (2009). The values of P100" are determined from the other columns according to (8).

| $u_i$ | $n_i$ | $i$ | P100 | $j$ | P100' | $N_i$ | percentiles InCites | PiIC | PRou | PPaG | P100" |
|---|---|---|---|---|---|---|---|---|---|---|---|
| 69 | 1 | 33 | 100.0 | 3423 | 100.00 | 1 | 0.03 | 99.97 | 100.00 | 100.00 | 100.00 |
| 54 | 1 | 32 | 97.0 | 3422 | 99.97 | 2 | 0.06 | 99.94 | 99.97 | 99.97 | 99.97 |
| 52 | 1 | 31 | 93.9 | 3421 | 99.94 | 3 | 0.09 | 99.91 | 99.94 | 99.94 | 99.94 |
| 40 | 1 | 30 | 90.9 | 3420 | 99.91 | 4 | 0.12 | 99.88 | 99.91 | 99.91 | 99.91 |
| 32 | 1 | 29 | 87.9 | 3419 | 99.88 | 5 | 0.15 | 99.85 | 99.88 | 99.88 | 99.88 |
| 30 | 1 | 28 | 84.8 | 3418 | 99.85 | 6 | 0.18 | 99.82 | 99.85 | 99.85 | 99.85 |
| 28 | 1 | 27 | 81.8 | 3417 | 99.82 | 7 | 0.20 | 99.80 | 99.82 | 99.82 | 99.82 |
| 27 | 1 | 26 | 78.8 | 3416 | 99.80 | 8 | 0.23 | 99.77 | 99.80 | 99.80 | 99.79 |
| 25 | 1 | 25 | 75.8 | 3415 | 99.77 | 9 | 0.26 | 99.74 | 99.77 | 99.77 | 99.76 |
| 24 | 2 | 24 | 72.7 | 3413 | 99.71 | 11 | 0.32 | 99.68 | 99.74 | 99.72 | 99.72 |
| 23 | 2 | 23 | 69.7 | 3411 | 99.65 | 13 | 0.38 | 99.62 | 99.68 | 99.66 | 99.66 |
| 22 | 2 | 22 | 66.7 | 3409 | 99.59 | 15 | 0.44 | 99.56 | 99.62 | 99.61 | 99.60 |
| 21 | 1 | 21 | 63.6 | 3408 | 99.56 | 16 | 0.47 | 99.53 | 99.56 | 99.56 | 99.55 |
| 20 | 1 | 20 | 60.6 | 3407 | 99.53 | 17 | 0.50 | 99.50 | 99.53 | 99.53 | 99.52 |
| 19 | 2 | 19 | 57.6 | 3405 | 99.47 | 19 | 0.55 | 99.45 | 99.50 | 99.49 | 99.48 |
| 18 | 3 | 18 | 54.5 | 3402 | 99.39 | 22 | 0.64 | 99.36 | 99.45 | 99.42 | 99.41 |
| 17 | 5 | 17 | 51.5 | 3397 | 99.24 | 27 | 0.79 | 99.21 | 99.36 | 99.30 | 99.29 |
| 16 | 3 | 16 | 48.5 | 3394 | 99.15 | 30 | 0.88 | 99.12 | 99.21 | 99.18 | 99.17 |
| 15 | 7 | 15 | 45.5 | 3387 | 98.95 | 37 | 1.08 | 98.92 | 99.12 | 99.04 | 99.01 |
| 14 | 9 | 14 | 42.4 | 3378 | 98.69 | 46 | 1.34 | 98.66 | 98.92 | 98.80 | 98.77 |
| 13 | 8 | 13 | 39.4 | 3370 | 98.45 | 54 | 1.58 | 98.42 | 98.66 | 98.55 | 98.51 |
| 12 | 6 | 12 | 36.4 | 3364 | 98.28 | 60 | 1.75 | 98.25 | 98.42 | 98.35 | 98.31 |
| 11 | 11 | 11 | 33.3 | 3353 | 97.96 | 71 | 2.07 | 97.93 | 98.25 | 98.10 | 98.03 |
| 10 | 20 | 10 | 30.3 | 3333 | 97.37 | 91 | 2.66 | 97.34 | 97.93 | 97.65 | 97.52 |
| 9 | 36 | 9 | 27.3 | 3297 | 96.32 | 127 | 3.71 | 96.29 | 97.34 | 96.83 | 96.58 |
| 8 | 35 | 8 | 24.2 | 3262 | 95.30 | 162 | 4.73 | 95.27 | 96.29 | 95.79 | 95.52 |
| 7 | 46 | 7 | 21.2 | 3216 | 93.95 | 208 | 6.07 | 93.93 | 95.27 | 94.61 | 94.21 |
| 6 | 79 | 6 | 18.2 | 3137 | 91.64 | 287 | 8.38 | 91.62 | 93.93 | 92.79 | 92.04 |
| 5 | 94 | 5 | 15.2 | 3043 | 88.90 | 381 | 11.13 | 88.87 | 91.62 | 90.26 | 89.29 |
| 4 | 167 | 4 | 12.1 | 2876 | 84.02 | 548 | 16.00 | 84.00 | 88.87 | 86.45 | 84.59 |
| 3 | 251 | 3 | 9.1 | 2625 | 76.69 | 799 | 23.34 | 76.66 | 84.00 | 80.34 | 77.33 |
| 2 | 405 | 2 | 6.1 | 2220 | 64.86 | 1204 | 35.16 | 64.84 | 76.66 | 70.77 | 65.55 |
| 1 | 670 | 1 | 3.0 | 1550 | 45.28 | 1874 | 54.73 | 45.27 | 64.84 | 55.07 | 45.86 |
| 0 | 1550 | 0 | 0.0 | 0 | 0.00 | 3424 | 100.00 | 0.00 | 45.27 | 22.65 | 0.00 |

**Table 2.** For a small model dataset, the values of the indicators P100 and P100' are compared with the uncertainty interval (Leydesdorff, 2012). There are 4 unique citation counts which lead to the ranks *i* which yield the P100 values. The numbers $n_i$ of papers determine the P100' indicator. The lower boundaries of the uncertainty intervals are given by the inverted InCites scale, the upper boundaries by the rule of Rousseau (2012). The initial situation is modified by shifting one of the uncited papers to the first and to the second unique citation count.

| | | initial situation | | | first modification | | | second modification | | |
|---|---|---|---|---|---|---|---|---|---|---|
| *i* | P100 | $n_i$ | P100' | unc. int. | $n_i$ | P100' | unc. int. | $n_i$ | P100' | unc. int. |
| 0 | 0 | 2 | 0 | 0 – 40 | 1 | 0 | 0 – 20 | 1 | 0 | 0 – 20 |
| 1 | 33 | 1 | 50 | 40 – 60 | 2 | 25 | 20 – 60 | 1 | 25 | 20 – 40 |
| 2 | 67 | 1 | 75 | 60 – 80 | 1 | 75 | 60 – 80 | 2 | 50 | 40 – 80 |
| 3 | 100 | 1 | 100 | 80 – 100 | 1 | 100 | 80 – 100 | 1 | 100 | 80 – 100 |

**Table 3.** Same as Table 2, but for increasing number $n_3$ of most cited papers.

| | | third modification | | | forth modification | | | fifth modification | | |
|---|---|---|---|---|---|---|---|---|---|---|
| *i* | P100 | $n_i$ | P100' | unc. int. | $n_i$ | P100' | unc. int. | $n_i$ | P100' | unc. int. |
| 0 | 0 | 1 | 0 | 0 – 20 | 1 | 0 | 0 – 17 | 1 | 0 | 0 – 14 |
| 1 | 33 | 1 | 33 | 20 – 40 | 1 | 33 | 17 – 33 | 1 | 33 | 14 – 29 |
| 2 | 67 | 1 | 67 | 40 – 60 | 1 | 67 | 33 – 50 | 1 | 67 | 29 – 43 |
| 3 | 100 | 2 | 100 | 60 – 100 | 3 | 100 | 50 – 100 | 4 | 100 | 43 – 100 |

**Table 4.** Same as Table 3, but for 5 unique citation counts. The initial situation is modified, so that the third modification in Table 3 is obtained.

| initial situation | | | first modification | | |
|---|---|---|---|---|---|
| *i* | $n_i$ | P100' | *i* | $n_i$ | P100' |
| 0 | 1 | 0 | 0 | 1 | 0 |
| 1 | 1 | 25 | 1 | 1 | 33 |
| 2 | 1 | 50 | 2 | 1 | 67 |
| 3 | 1 | 75 | | | |
| 4 | 1 | 100 | 3 | 2 | 100 |

**Table 5.** The values of the new indicator P100" for the initial situation and the 5 modifications from Tables 2 and 3.

| | modification | | | | | |
|---|---|---|---|---|---|---|
| *i* | 0 | 1 | 2 | 3 | 4 | 5 |
| 0 | 0 | 0 | 0 | 0 | 0 | 0 |
| 1 | 47 | 34 | 27 | 27 | 23 | 19 |
| 2 | 74 | 74 | 67 | 54 | 44 | 39 |
| 3 | 100 | 100 | 100 | 100 | 100 | 100 |

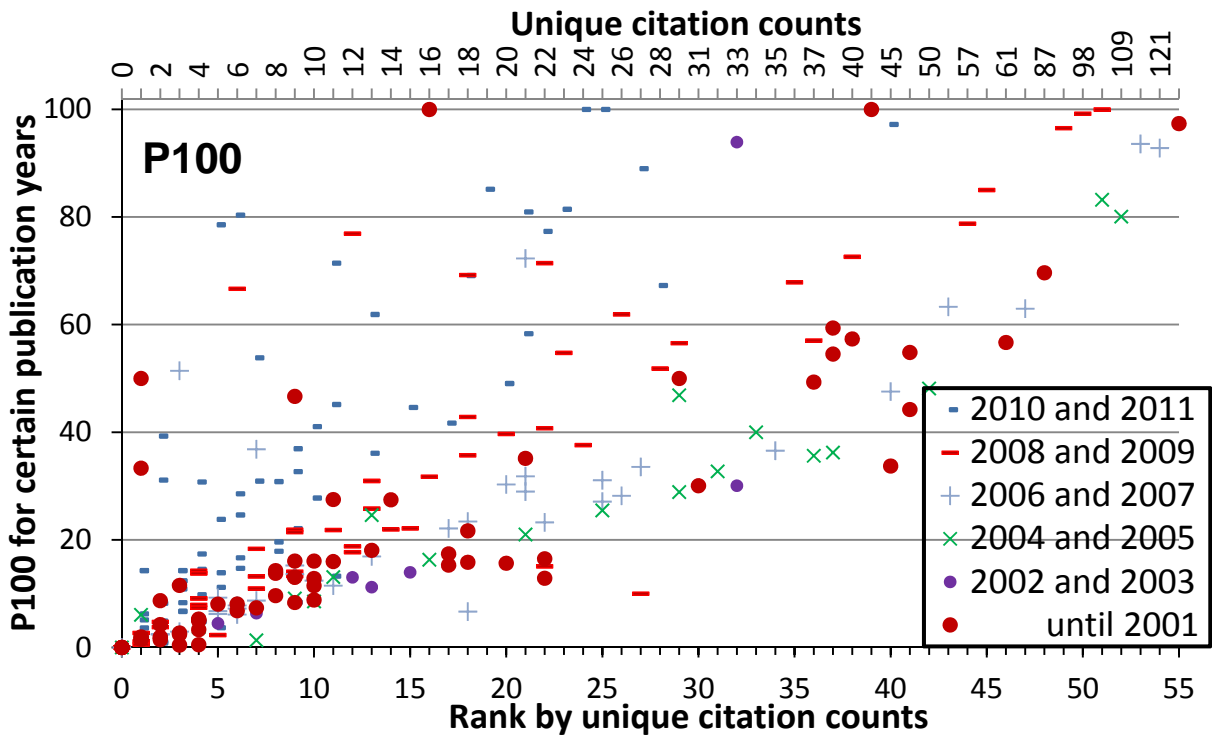

**Fig. 1.** Values of the P100 indicator for the articles of the three senior authors who have proposed the indicator. Different publication periods are distinguished. Only citation counts for which at least one article is published by one of the authors are utilized. These are the same data as in Fig. 1 of my previous investigation (Schreiber, 2014) but on a linear scale.

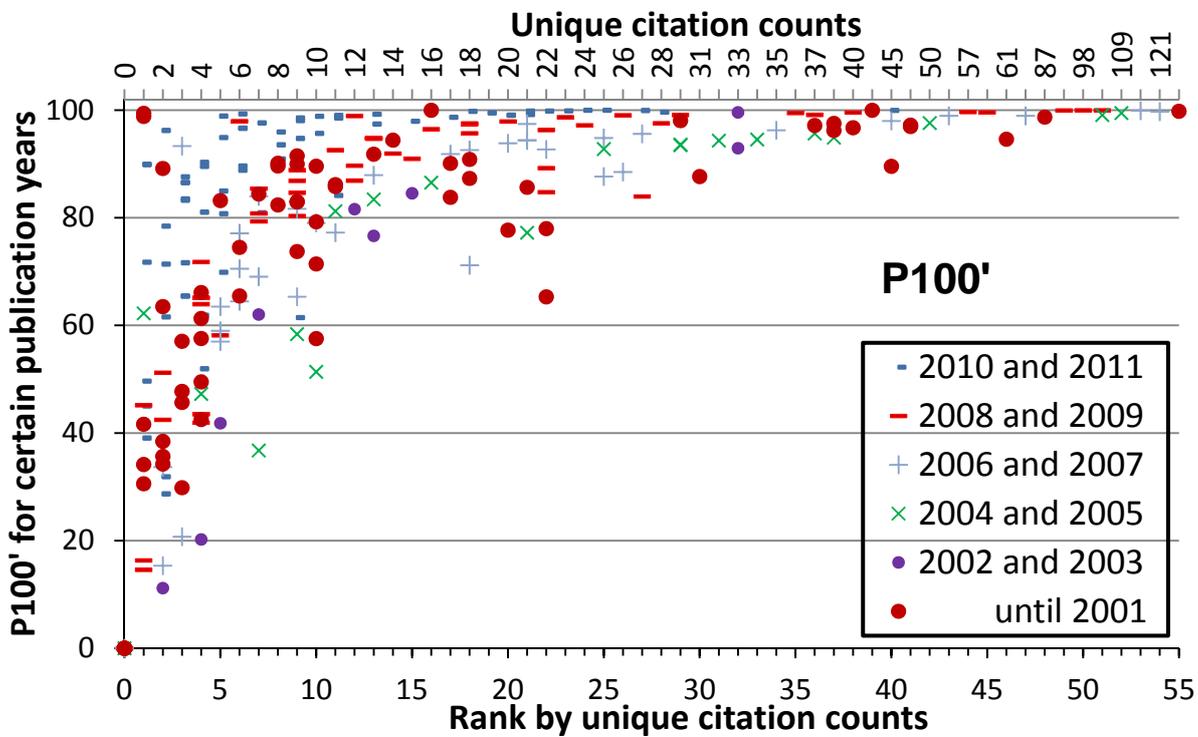

**Fig. 2.** Same as Fig. 1, but for the indicator P100'.

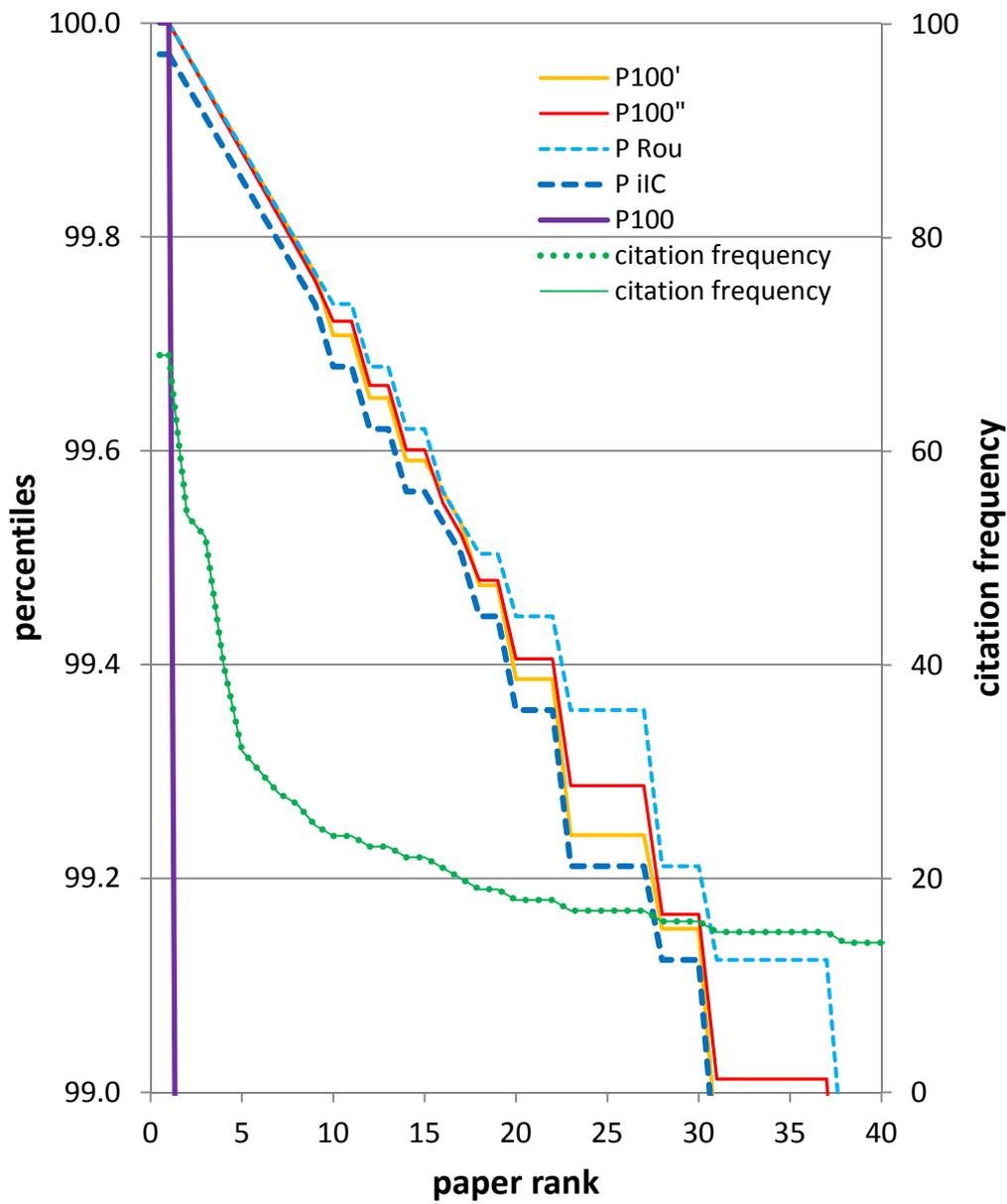

**Fig. 3.** Different percentile values in comparison with the indicators P100, P100', and P100". The sequence from top to bottom is determined by PRou ≥ P100" ≥ P100' ≥ PiIC ≥ P100. The papers are ranked by decreasing number of citations. The citation frequency is also displayed (lowest thin curve with dots, right hand scale). The top and the bottom of the uncertainty interval are given by the thin and thick broken lines. For the uncited papers all values except PRou are equal to zero and cannot be distinguished in the plot. For higher citation frequencies P100' and PiIC are so close that they also cannot be distinguished on this scale.

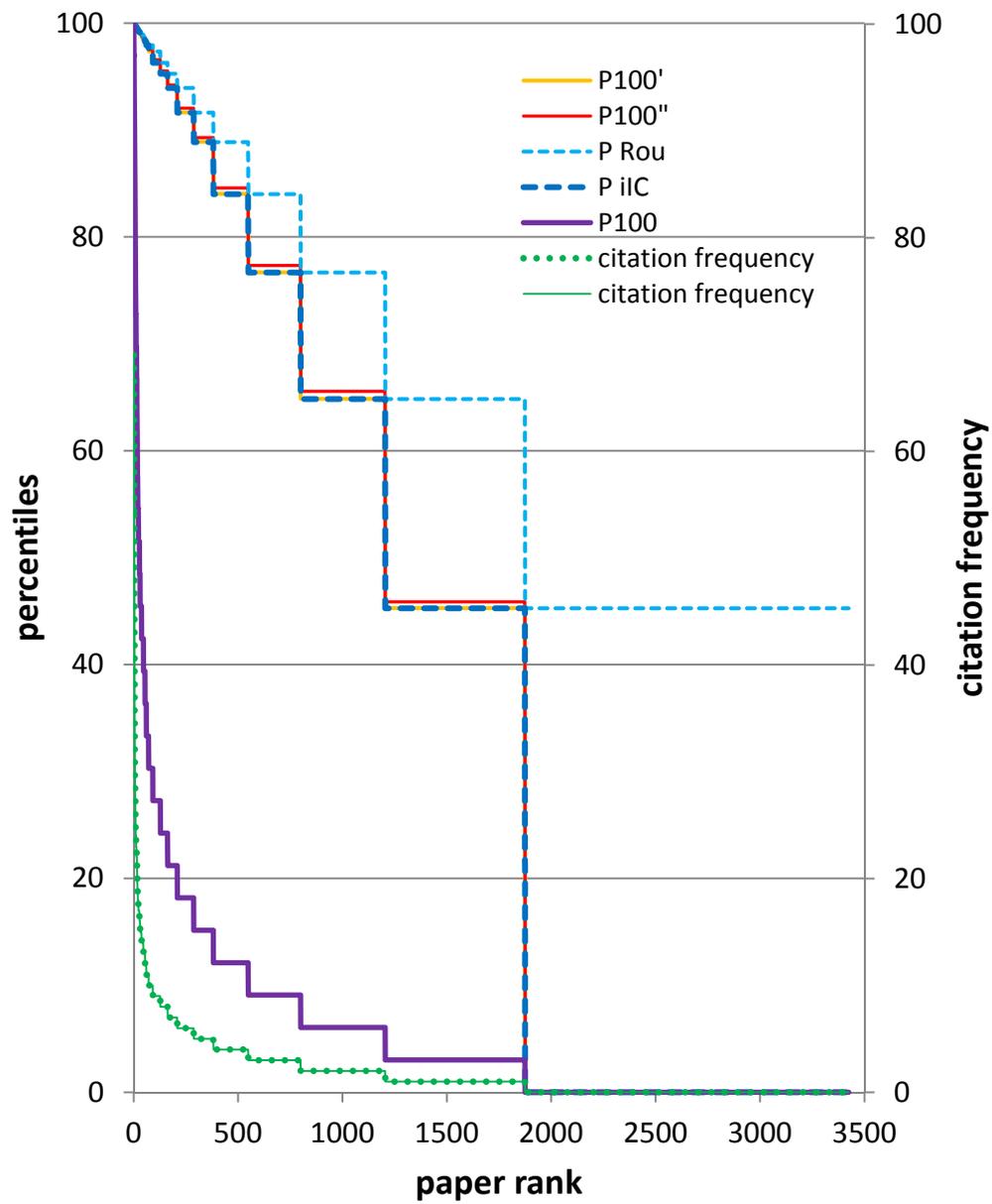

**Fig. 4.** Same as Fig. 3, but enlarged. Now a distinction between the different indicators is possible. P100" lies usually above P100', except for those cases when the width $n_i/N$ of the uncertainty interval equals $1/N$.